\documentstyle[12pt,iopfts]{ioplppt}
\def\rf#1{(\ref{#1})}

\def\tgam{\tilde{\gamma}}
\def\ksi#1{\stack{#1}{x}}
\def\bbox#1{{\bi #1}}
\def\myrm{\vphantom{i}}

\newcommand{\stack}[2]
 {\stackrel{\scriptstyle #1}{#2}\hspace{-3.5pt}\vphantom{#2}}

\begin{document}

\title{ Quantization of fields over de Sitter space  \\
 by the method of generalized coherent states. \\
 II. Spinor field}[Quantization by the generalized coherent states.II]
\author{S A Pol'shin\ftnote{1}{E-mail: itl593@online.kharkov.ua}}
\address{Department of Physics, Kharkov State University, \\
  Svobody Sq., 4, 310077, Kharkov, Ukraine }
\date{}
\jl{1}

\begin{abstract}
Connection of the invariant Dirac equation over the de Sitter space with
irreducible representations of the de Sitter group is ascertained. The set of
solutions of this equation is obtained in the form of the product of two
different systems of generalized coherent states for the de Sitter group.
Using  these solutions the quantized Dirac field over de Sitter space
is constructed and its propagator is found. It is
 a result of action of some de Sitter invariant spinor operator onto the
spin zero propagator with an imaginary shift of a mass.
\end{abstract}

\section{Introduction}

It seems that in the case of spin~1/2 particles the usual methods
are insufficient for consistent construction of a theory of quantized field
over the de Sitter (dS) space. Indeed, a lot of papers were concerned with
obtaining the solutions of covariant (\cite{16/65} and references therein)
and group theoretical~\cite{87} Dirac equation over the dS space by the
method of separation of variables, however all these solutions have
complicated form which considerably troubles the construction of the theory
of quantized field.  Only in the little-known paper~\cite{80} the summation
over one of such a set of solutions was performed; the resulting
propagator is not dS-invariant and does not obey the causality principle. In
the over hand, in~\cite{61} a spinor propagator was found starting from the
demands of dS-invariant Dirac equation satisfaction, dS-invariance and the
boundary conditions; but the quantized field to which it  corresponds was not
found; on the contrary, in the Anti-de Sitter space the quantized spinor
fields with an invariant propagator was constructed long ago~\cite{36}.

In the present paper we show that all these troubles may be overcame using
the method of generalized coherent states (CS), and build the theory of
quantized spinor field over the dS space by analogy with the theory of scalar
field considered in the part~I of this series of paper~\cite{dS1}.
The present paper is composed as follows.
In section 2 we consider the dS-invariant Dirac
equation and show that the corresponding representation of the dS group is
irreducible and falls under the classification listed in section 3 of part~I.
Also we show that this equation admits the reduction to the covariant
form by the well simpler way than proposed previously~\cite{61}. In section 3
we construct the CS system for the four-spinor representation of the dS group
in the form of $4\times 2$-matrices. Solutions of the dS-invariant Dirac
equation are the products of these CS and scalar CS obtained in section 4 of
part~I. In fact, these solutions are the more compact form of the spinor
"plane waves" obtained in~\cite{dS-PLB}. The invariance properties of these
solutions allow us to construct of them a dS-invariant two-point function
and compute it passing to the complexified dS space. In section 4
 we construct the quantized spinor field using these solutions; its
propagator is expressed  by the boundary values of two-point functions
obtained in section 3 and coincides with the expression obtained {\it a
priori} in~\cite{61}, to within the constant multiplier. In section 5 we
briefly summarize the results of parts~I and~II of this series of paper.

\section{The Dirac equation}

Introducing the matrices
\[\gamma ^{5}={\i}\gamma ^{0}\gamma ^{1}\gamma ^{2}\gamma ^{3} \qquad
\tgam^{\mu} =-{\i}\gamma ^{5}\gamma ^{\mu} \qquad
 \tgam^{5}=\i\gamma ^{5} \]
let us write down the generators of four-spinor representation of the
dS group in the five-dimensional form:
\begin{equation}\label{5. 39a}
J^{(s)AB}=\frac{1}{4} [\tgam ^{A}, \tgam ^{B}].
\end{equation}
The equalities
\begin{equation}\label{5.  40a}
 \tgam^{A}\tgam^{B}+
\tgam^{B}\tgam^{A}=2\eta ^{AB}
\end{equation}
\begin{equation}
\label{5.  40b}
\tgam^{A}\tgam^{B}\tgam^{C}=
\eta ^{AB}\tgam^{C}+\eta ^{BC}\tgam^{A}-
\eta ^{AC}\tgam^{B}+\frac{1}{2}
\varepsilon ^{ABCDE}\tgam_{D}\tgam_{E}
\end{equation}
\begin{equation}
\label{5.  40c}
\eqalign{
\tgam^{A}\tgam^{B}\tgam^{C}\tgam^{D}
=\eta ^{AB}\tgam^{C}\tgam^{D}+
\eta ^{BC}\tgam^{A}\tgam^{D}-
\eta ^{AC}\tgam^{B}\tgam^{D}+ \\
+2(\eta ^{AD}J^{(s)BC}+\eta ^{CD}J^{(s)AB}-\eta ^{BD}J^{(s)AC} )-
\varepsilon ^{ABCDE}\tgam_{E}
}
\end{equation}
hold. With the help of Equations (8) and (9) of Part~I
\vphantom{ref{5.  26})ref{5.  28}}
and~\rf{5.  40a}-(\ref{5. 40c}) we obtain
\begin{equation}\label{5.  41}
 R^{2}C^{(s)}_{2}=5/2  \qquad
\label{5.  42}
 W^{(s)}_{A}=\frac{3}{4}\tgam_{A} \qquad
\label{5.  43}
  R^{2}C^{(s)}_{4}=\frac{45}{16}.
  \end{equation}
Comparing the above expression
with Equation (10) of Part~I\vphantom{ref{5.  29})ref{5.  30}}
we see that it is the representation $\bpi_{3/2, 3/2}$.
We shall choose the standard form of $\gamma$-matrices. Then it is easy to
show that the explicit form of generators is
\begin{eqnarray}\label{3.  11a}
\bPi^{+(s)}=\frac{1}{R}\left(
\begin{array}{rr}
0 & -\bsigma \\
0 & 0
\end{array}
\right) \qquad
\bPi^{-(s)}=\frac{1}{R}\left(
\begin{array}{rr}
0 & 0 \\
\bsigma & 0
\end{array}
\right) \\
\label{3.  11c}
P^{0(s)}=\frac{1}{2}\left(
\begin{array}{rr}
1 & 0 \\
0 & -1
\end{array}
\right) \qquad
J_{ik}^{(s)}=-\i\varepsilon_{ikl}
\left(
\begin{array}{ll}
\sigma^{l} & 0 \\
0 & \sigma^{l}
\end{array}
\right) . \nonumber
\end{eqnarray}
We denote matrices of finite transformations  as $U(g)$.
For their were meeting
 the same composition properties as the coordinate
transformations, it is necessary to take all the parameters
of  transformations with
the sign which is opposite with respect to the scalar representation.
Then we obtain
\begin{eqnarray}\label{3.  13a}
 U(\Theta_{\pm}({{\bi a}}))=1-\bPi^{\pm (s)}{{\bi a}}R \nonumber \\
\label{3.  13b}
U(\Theta_{0}(\varepsilon))=\exp (-P^{(s)}_{0}\varepsilon). \nonumber
\end{eqnarray}

Let us consider the representation
$\bpi_{3/2, 3/2}\otimes \bnu_{m, 0}$.
Its generators are the sum of generators~(13) of Part~I
\vphantom{ref{5. 34}} ({\it orbital}
part) and generators~(\ref{5. 39a}) ({\it spin} part).
Then the second-order Casimir operator is equal to
\[ C_{2}=C^{(l)}_{2}+C^{(s)}_{2}-R^{-2}J^{(s)AB}J^{(l)}_{AB}.  \]
Denoting
$\hat{\nabla}_{\rm dS}=-R^{-1}J^{(s)AB}J^{(l)}_{AB}$
we obtain
\begin{equation}\label{5.  45}
C_{2}=\Box  +\frac{\hat{\nabla}_{\rm dS}}{R}+\frac{5}{2R^{2}}.
\end{equation}
To compute the fourth-order Casimir operator we write according to
 Equation (14) of Part~I\vphantom{ref{5.  35}} and~\rf{5.  42}:
\[ RW_{A}=
-\frac{1}{8}\varepsilon _{ABCDE}\tgam^{B} \tgam^{C}
J^{(l)DE}+\frac{3}{4}\tgam_{A}.  \]
Squaring $W_{A}$  it is necessary to use
formulas~\rf{5.  40a}-(\ref{5. 40c}). The result obtained
\[ C_{4}=\frac{3}{4}\Box+\frac{3\hat{\nabla}_{\rm
dS}}{4R}+\frac{45}{16R^{2}}\]
is in agreement with Equation (12) of Part~I
\vphantom{ref{5.  32}}  at $s=1/2$ and~\rf{5.  45}.
From the second Shur's lemma  follows that the operators
$\hat{\nabla}_{\rm dS}$  and $\Box$
should have fixed eigenvalues in the irreducible representations.
Then using Equation (11) of Part~I
\vphantom{ref{5.  31}},~\rf{5.  45} and the equality
\[\hat{\nabla}_{\rm dS}^{2}=\frac{1}{4R^{2}}\tgam^{A}\tgam^{B}\tgam^{C}
\tgam^{D}J^{(l)}_{AB}J^{(l)}_{CD}=\Box -3\hat{\nabla}_{\rm dS}/R\]
we obtain the quadratic equation for eigenvalues of
$\hat{\nabla}_{\rm dS}$. Solving it yields
\begin{eqnarray}\label{5. 46a}
\hat{\nabla}_{\rm dS}=-2R^{-1}\pm {\i}\mu \\
\Box =-\mu ^{2}\mp \i R^{-1}\mu -2R^{-2} \nonumber
\end{eqnarray}
where $\mu ^{2}=m^{2}-\frac{1}{4R^{2}}.$ As
$m^{2}>1/4R^{2}$ (see section 3 of part~I),
then $\mu$ is a real number. The appearance of two signs
indicates that two identical irreducible representations had appeared:
\[\bnu_{m, 0} \otimes \bpi_{3/2, 3/2}=2\bnu_{m, 1/2}.\]
Using Equation (13) of Part~I
\vphantom{ref{5.  34}}  we can write
\[\hat{\nabla}_{\rm dS}=\Gamma ^{\mu}\partial _{\mu} \qquad
 \Gamma ^{\mu}=\chi  \gamma ^{\mu}
+\frac{1}{2R} [\gamma ^{\mu}, \gamma _{\nu}]x^{\nu}.\]
Choosing the representation which corresponds to the lower sign
in~(\ref{5. 46a}) we finally obtain the group theoretical Dirac
equation over the dS space:
\begin{equation}\label{5.  47}
{\i}\Gamma ^{\mu}\partial _{\mu}\psi
-(\mu -\frac{2\i}{R})\psi =0.
\end{equation}
This was well known previously out of the context of dS group irreducible
representations~\cite{Gursey}.

The above equation
 admits the transformation into the
covariant form. To this end let us perform the transformation
$\Psi =V\psi,$ where
\[\label{5.  47a}
V=(1-\varepsilon_{\mu}\varepsilon^{\mu} )^{-1/2}
(1+\gamma_{\mu}\varepsilon^{\mu} ) \qquad
\varepsilon ^{\mu}=\frac{x^{\mu}}{R(\chi +1)}.\]
Then~\rf{5. 47} passes into
 \begin{equation}\label{5.  48}
 \i V\Gamma ^{\mu}V^{-1}(\partial _{\mu}\Psi
 +(V\partial _{\mu}V^{-1})\Psi )-(\mu -2\i R^{-1})\Psi =0.
\end{equation}
It is easy to show that
\begin{equation}\label{5.  49}
\eqalign{
 V\Gamma ^{\mu}V^{-1}=e^{\mu}_{(\nu )}\gamma ^{\nu} \\
\label{5.  50}
  \partial _{\mu}+V\partial _{\mu}V^{-1} ={\cal D}_{\mu}-
\frac{1}{2R}\gamma _{\mu}
}
\end{equation}
where $e^{(\mu)}_{\nu}$ is the vierbein which is orthonormal with respect to
the metric~(3) of Part~I\vphantom{ref{5. 21}}:
 \[\label{5.  51}
e^{(\mu) \nu}=\eta ^{\mu \nu}+\frac{x^{\mu}x^{\nu}}{R^{2}(\chi +1)}\]
and  ${\cal D}_{\mu}$ is the spinor covariant derivative
\[\fl {\cal D}_{(\mu)}=e_{(\mu)}^\nu
\partial_{\nu}-\frac{1}{2}J^{(s)\nu\rho} G_{\nu\rho\mu} \qquad
G_{\nu\rho\mu}=
e_{(\nu);\kappa}^\sigma e_{(\rho)\sigma}e^\kappa_{(\mu)}
=\frac{1}{R^{2}(\chi +1)}(x_{\nu}\eta_{\mu \rho}-
x_{\rho}\eta_{\nu\mu}).  \]
Then putting together~(\ref{5. 48})-(\ref{5. 50}) we finally obtain
\[\i\gamma ^{\mu} e^{\nu}_{(\mu )} {\cal D}_{\nu}\Psi =\mu \Psi. \]
Another more complicated way of transformation of dS-invariant
Dirac equation to the covariant one was proposed in~\cite{61}.

\section{Spinor coherent states}
Let us denote the constant $4\times 2$-matrices as
$A,A',A''$ and  define over such  matrices the weak equivalence
relation $\sim$ and the strong one $\simeq$ as
\begin{eqnarray}
A'\sim A'' \Leftrightarrow A'= A''B \qquad  B\in GL(2,{\Bbb C})
\nonumber \\
A'\simeq A'' \Leftrightarrow A'= A''B \qquad  B\in SU(2). \label{3. 35}
\nonumber
\end{eqnarray}
Also, let us define the product of two $4\times 2$-matrices
$A'$ and $A''$ as $A'\overline{A''}$, where the upper line
denotes the Dirac conjugation. Consider the left action of four-spinor
representation of the dS group over these matrices:
 $g:\ A\mapsto U(g)A.$ It is easy to show that the matrices
\[\label{vectors+-}
|+\rangle=\left(
\begin{array}{l}
I_2 \\ 0_2
\end{array}
\right)   \qquad
|-\rangle=\left(
\begin{array}{l}
0_2 \\ I_2
\end{array}
\right)\]
(where $I_2$ is the unit $2\times 2$-matrix) are invariant
under transformations which belong to the subgroups
${\cal K}^\pm \equiv {\cal T}^\pm \circledS ({\cal T}^0 \otimes {\cal R})$
to within the weak equivalence relation.
In the terms of the strong equivalence relation we have
\[ U(h)|\pm\rangle \simeq (\alpha^\pm_{\bbox{v}}(h))^{-1/2}
|_{\bbox{v}=\bbox{0}}|\pm\rangle \qquad  h\in {\cal K}^\pm .\]
From the other hand, it is easily seen that the subgroups
${\cal K}^\pm$ are stability subgroups of the vector
$\bbox{w}=\bbox{0}$ concerning the conformal action~(19) of Part~I
\vphantom{ref{3. 15}}
of the dS group. This allows us to identify the
$SO(4,1)/{\cal K}^\pm$ space with the space ${\Bbb R}^3$ of vectors
$\bbox{w}$. As the lifting from the
$SO(4,1)/{\cal K}^\pm$ space to the dS group we shall take the transformation
which transforms the origin into the point ${\bi w}$:
\[ SO(4,1)/{\cal K}^\pm \ni \bbox{w}\mapsto
g_{\bbox{w}}=\Theta_\mp (-\bbox{w}) \in SO(4,1).\]
Then the CS system for the $SO(4,1)/{\cal K}^\pm$ space
being dS-invariant to within the weak equivalence relation is
\begin{eqnarray}
|\bbox{w}\pm\rangle =U(g_{\bbox{w}})|\pm\rangle \nonumber \\
|\bbox{w}+\rangle=\left(
\begin{array}{l}
I_2 \\ \bbox{\sigma}\bbox{w}
\end{array}
\right)   \qquad
|\bbox{w}-\rangle=\left(
\begin{array}{l}
-\bbox{\sigma}\bbox{w} \\ I_2
\end{array}
\right). \nonumber
\end{eqnarray}
With the help of Equation (1) of Part~I
\vphantom{ref{foliation}} the transformation properties of
these vectors  with respect to the strong equivalence relation
may be written as
\begin{equation}\label{transf-dS-1/2}
 U(g_1 )|\bbox{w}\pm\rangle \simeq
(\alpha^\pm_{\bbox{v}}(g_{\bbox{w}'}^{-1}g_1 g_{\bbox{w}})
)^{-1/2}|_{\bbox{v}=\bbox{0}}
|{\bbox{w}}_{g_1} \pm\rangle \qquad  g\in {\cal G}.
\end{equation}
As the transformations $T^\pm_\sigma (g)$ compose a representation of the dS
group then
\[\alpha^\pm_{\bbox{v}}(g_2 g_1 )=\alpha^\pm_{\bbox{v}}(g_2 )
\alpha^\pm_{\bbox{v}'}(g_1) \qquad  g_1 ,g_2 \in {\cal G}
\qquad  \bbox{v}'=\bbox{v}_{g_2^{-1}}.\]
Then using the above expression and Equation (1) of Part~I
\vphantom{ref{foliation}} we get
\[\alpha^\pm_{\bbox{v}}(g_1)=\alpha^\pm_{\bbox{v}'}
(g_{\bbox{w}'}^{-1}g_1 g_{\bbox{w}})  \qquad
\bbox{v}'=\bbox{v}_{g_{\bbox{w}'}^{-1}} \qquad
\bbox{w}'=\bbox{w}_{g_1} .\]
Putting $\bbox{v}=\bbox{w}'$ in the above expression, we can rewrite the
transformation properties~(\ref{transf-dS-1/2}) as
\begin{equation}\label{3. 37}
(\alpha^\pm_{\bbox{w}}(g))^{1/2}U(g)|\bbox{w}_{g^{-1}}\pm\rangle
\simeq |\bbox{w}\pm\rangle \qquad  g\in {\cal G}.
\end{equation}
It is easy to show that the equalities
\begin{eqnarray}\label{3. 36}
\left(\gamma\cdot k_{{\bi w}}\mp 1 \right)|{{\bi w}}\pm\rangle =0 \\
|{\bi w}\pm\rangle\langle {\bi w}\pm | =\frac{1-{\rm w}^2}{2}
(\gamma\cdot k_{{\bi w}}\pm 1) \label{<w|w>}
\end{eqnarray}
are correct. Now let us construct  the $4\times 2$-matrix functions
\[\Phi_{{\bi w}}^{(1/2)\pm}(x)=
\Phi_{{\bi w}}^{(0)\pm}(x;\sigma_0 -1/2)|{{\bi w}}\pm\rangle.\]
Using~(\ref{3. 36}) we obtain that they obey~(\ref{5. 47}):
\[ ( \i\hat{\nabla} -\mu+2\i R^{-1}) \Phi_{\bbox{w}}^{(1/2)\pm}(x)=0.\]
These solutions are much simpler that these obtained by the method of
separation of variables~\cite{16/65,87}.

From the transformation properties~(21) of Part~I\vphantom{ref{transf}}
and~(\ref{3. 37}) it follows that under  transformations from the dS group
the functions $\Phi_{{\bi w}}^{(1/2)\pm}(x)$ transform just as the functions
$\Phi_{{\bi w}}^{(0)\pm}(x;\sigma_0)$, to within the constant matrix
transformation:
\begin{equation}\label{3. 27}
\Phi_{{\bi w}}^{(1/2)\pm}(x_{g})\simeq
(\alpha^{\pm}_{{\bi w}}(g))^{\sigma_0} U(g)
\Phi_{{{\bi w}}'}^{(1/2)\pm}(x)
\end{equation}
where ${{\bi w}}'={{\bi w}}_{g^{-1}}$.
As the inversion $\bbox{w}\mapsto -\bbox{w}/{\myrm w}^2$ yields
\[\Phi^{(1/2)\pm}_{-\bbox{w}/{\myrm w}^2}(x) \simeq
-i (-{\myrm w}^2)^{-\sigma_0}\Phi^{(1/2)\mp}_{\bbox{w}}(x)\]
then the functions $\Phi_{{\bi w}}^{(1/2)\pm}(x)$ and
$\Phi_{{\bi w}}^{(1/2)\pm}(x)$ yield the
same two-point function. Let us define its as follows:
\[ \fl \frac{1}{8}{\cal W}^{(1/2)} (\ksi{1},\ksi{2})=
\int_{{\Bbb R}^3}\d^3 {{\bi w}}\,\Phi^{(1/2)+}_{{\bi w}}(\ksi{1})
\overline{\Phi}\vphantom{\Phi}^{(1/2)+}_{{\bi w}}(\ksi{2})
\lo= \int_{{\Bbb R}^3}\d^3 {{\bi w}}\,\Phi^{(1/2)-}_{{\bi w}}(\ksi{1})
\overline{\Phi}\vphantom{\Phi}^{(1/2)-}_{{\bi w}}(\ksi{2}). \]
From~(\ref{3. 27})  follows that the above functions are dS-invariant in the
sense that at $g\in\cal G$
\[\label{3.  33}
{\cal W}^{(1/2)}(\ksi{1}_{g},\ksi{2}_{g})=
U(g){\cal W}^{(1/2)}(\ksi{1},\ksi{2})\overline{U}(g).\]
Using~(\ref{<w|w>})  it is easy to show that
\[ \fl {\cal W}^{(1/2)}(\ksi{1},\ksi{2})
= \frac{1}{2}\int_{S^3}\frac{\d^3{\bi l}}{l^5}
\left( \frac{\stack{1}{x}^0+l^a\stack{1}{x}^a}{R}
\right)^{-\i\mu R-2}
\left( \frac{\stack{2}{x}^0+l^a \stack{2}{x}^a}{R}
\right)^{\i\mu R-2}
(\gamma^0 +\bbox{\gamma}{\bi l}+l^5 ). \label{dirac-3-sphere}\]
As the functions $\varphi_{k}^{(1/2)\pm}(\zeta;\lambda)$ inherit the
analyticity properties of functions $\varphi_{k}^{(0)\pm}(\zeta;\sigma_0)$
over the complexified dS space, then, as in the spin zero case, the
functions ${\cal W}^{(1/2)}(\stack{1}{x},\stack{2}{x})$ converge at
$(\stack{1}{\zeta},\stack{2}{\zeta})\in {\cal D}^+ \times {\cal D}^- .$
Choosing the points according
to Equation (23) of Part~I\vphantom{ref{points-dS}} and using the equality
\begin{eqnarray}
\fl\int_{0}^\pi \d\theta \, \sin^2 \theta \cos\theta
(\cosh v+\sinh v\cos\theta )^{-\i\mu R-2}  \nonumber\\
\lo=\frac{\pi}{\sinh v}\left(
\mathop{_2 F_1}
\left( 1-\frac{i\mu R}{2}, \frac{1}{2}+\frac{i\mu R}{2};2; -\sinh^2 v\right)
\right. \nonumber \\ \left.
-\cosh v \mathop{_2 F_1}
\left( 1+\frac{i\mu R}{2}, \frac{1}{2}-\frac{i\mu R}{2};2; -\sinh^2 v\right)
\right) \nonumber
\end{eqnarray}
we obtain
\begin{equation}\label{twop-1/2-cdS}
\fl {\cal W}^{(1/2)}(\stack{1}{\zeta},\stack{2}{\zeta})
 = \frac{\pi^2\e^{-\pi\mu R}}{\mu -\i R^{-1}}
\tilde{\gamma}\vphantom{\gamma}_A
\stack{1}{\zeta}^A \left( \i\hat{\nabla}_{\rm dS}-\mu  +\frac{\i}{R}\right)
\mathop{_2 F_1}\left(2-\i\mu R,1+\i\mu R;2;\frac{1-\rho}{2}\right)\gamma^5
\end{equation}
where the operator $\hat{\nabla}_{\rm dS}$  acts onto the coordinates
$\stack{1}{\zeta}$.

\section{Spinor field over de Sitter space}

To construct a quantized spinor field, let us  use  the equality
\[ R^{-1}\tgam^A x_A
\left(\i\hat{\nabla}_{\rm dS} -\mu + 2\i R^{-1}\right) R^{-1} \tgam^B
x_B= \i\hat{\nabla}_{\rm dS} +\mu +2\i R^{-1}.\]
From here follows that
 if the function $\psi$ obey the Dirac equation~(\ref{5. 47}),
then the function $\tgam^A x_A \psi$ obey the same equation with the opposite
sign of $\mu$. Then the functions
\[\tilde{\Phi}\vphantom{\Phi}_{{\bi w}}^{(1/2)\pm}(x)=R^{-1}\tgam^A x_A
\Phi_{{\bi w}}^{(0)\pm}(x;\sigma_0^* -1/2)|{{\bi w}}\pm\rangle\]
obey Equation (\ref{5. 47}). Let us introduce also two sets of fermionic
creation-annihilation operators $b^{(\pm)}({{\bi w}})$
and $b^{(\pm)\dagger}({{\bi w}})$, which at the same time
 are  the matrices of dimensionality $2\times 1$ and $1\times 2$,
respectively, and obey the anticommutation relations
\begin{equation}\label{comm-fermionic}
\{ b^{(\pm)}({{\bi w}}) , b^{(\pm)\dagger}({{\bi w}}')\}
=\delta ({{\bi w}},{{\bi w}}')
\left(
\begin{array}{ll}
1 & 0 \\
0 & 1
\end{array}
\right)
\end{equation}
and all other anticommutators vanish.
Then we can construct the quantized spinor field as
\begin{equation}\label{field-dS-1/2}
\phi^{(1/2)}(x)=\int_{{\Bbb R}^3} \d^3 {{\bi w}}\,\left( \Phi^{(1/2)+}_{{\bi
w}} (x)b^{(+)}({{\bi w}}) + \tilde{\Phi}\vphantom{\Phi}_{{\bi w}}^{(1/2)-}(x)
b^{(-)}({{\bi w}}) \right).
\end{equation}
Using~(\ref{twop-1/2-cdS}) it is easy to show that the two-point function
which corresponds to the solutions
$\tilde{\Phi}\vphantom{\Phi}_{{\bi w}}^{(1/2)\pm}(x)$ is equal to
\begin{eqnarray}
\int_{{\Bbb R}^3}\d^3 {{\bi w}}\,
\tilde{\Phi}\vphantom{\Phi}_{{\bi w}}^{(1/2)\pm}(\stack{1}{\zeta})
\overline{\tilde{\Phi}}
\vphantom{\Phi}_{{\bi w}}^{(1/2)\pm}(\stack{2}{\zeta}) = \nonumber \\
\label{z2z1-z1z2}
R^{-2} \tgam^A \stack{1}{\zeta}_A
{\cal W}^{(1/2)} (\stack{2}{\zeta},\stack{1}{\zeta})
(\gamma^5 \tgam^B \stack{2}{\zeta}_B \gamma^5)=
-{\cal W}^{(1/2)} (\stack{1}{\zeta},\stack{2}{\zeta}).
\end{eqnarray}
Further, the hypergeometric functions in the r.h.s.
of~(\ref{twop-1/2-cdS}) differ from
${\cal W}^{(0)} (\stack{1}{\zeta},\stack{2}{\zeta})$ only by the constant
multiplier and the imaginary shift of mass. Then computing the
difference of its values on the edges of the cut
$z\in [1,+\infty)$ we can use the results of section 5 of part~I. Passing to
the boundary values Equation (\ref{z2z1-z1z2}) yields
\[ R^{-2} \tgam^A \stack{1}{x}_A{\cal W}^{(1/2)+}
 (\stack{2}{x},\stack{1}{x}) (\gamma^5 \tgam^B \stack{2}{x}_B \gamma^5)=
-{\cal W}^{(1/2)-} (\stack{1}{x},\stack{2}{x}) \]
which is analogous to Equation (27) of Part~I
\vphantom{ref{x1x2-x2x1}} for the spin zero case.
Then using Equation (28) of Part~I for the spin 1/2 propagator
\[\{ \phi^{(1/2)}_\alpha
(\ksi{1}),\overline{\phi}\vphantom{\phi}^{(1/2)}_\beta (\ksi{2}) \} \equiv
G^{(1/2)}_{\alpha\beta}(\ksi{1},\ksi{2})
={\cal W}^{(1/2)+}_{\alpha\beta} (\ksi{1},\ksi{2})-
{\cal W}^{(1/2)-}_{\alpha\beta}(\ksi{1},\ksi{2})\]
where $\alpha,\beta=1,\ldots,4$ are spinor indices, we finally obtain
\begin{eqnarray}
\fl G^{(1/2)}(\ksi{1},\ksi{2})
=\frac{\i\pi^2}{\mu -\i R^{-1}}(1+\e^{-2\pi\mu R}) \varepsilon
(\stack{1}{x}^0 -\stack{2}{x}^0) \tgam_A \stack{1}{x}^A
\nonumber \\ \times
\left(
\i\hat{\nabla}_{\rm dS}- \mu  +\frac{\i}{R}\right)\theta
\left(-\frac{1+G}{2}\right) \mathop{_2 F_1} \left(2-\i\mu R ,1+\i\mu R ;2;
\frac{1+G}{2}\right)\gamma^5 \nonumber
\end{eqnarray}
where the operator
$\hat{\nabla}_{\rm dS}$ acts onto the coordinates $\stack{1}{x}$. The above
expression coincides with the solution of Cauchy problem for the Dirac
equation over the dS space obtained in~\cite{61}. The only difference is that
our method do not allow us to find the behavior of propagator over the "light
cone" $G=-1$.

\section{Concluding remarks}

To summarize the results of the present series of papers,
we can say that the CS method allow us to quantize massive spin~0 and~1/2
fields over the dS space in the uniform way.
Both in the spin zero case and in the spin~1/2 one
the starting-point is the invariant wave equations which
correspond to  irreducible representations of the dS group. The solutions
of these equations are constructed from  CS  for the dS group; in the spin
zero case the dS-invariant Klein-Gordon equation is satisfied by the scalar
CS itself. In the spin~1/2 case the solutions of dS-invariant Dirac equation
are constructed from two different CS systems which correspond to
different representations of the dS group and different stationary subgroups.
Both in the spin zero case and in the spin~1/2 one these sets of solutions
possess the same transformation properties under the dS group, with
difference that the constant matrix transformation is added in the spin~1/2
case.

From these sets of solutions we can construct the two-point functions
${\cal W}^{(s)}(\stack{1}{x},\stack{2}{x})$ which have the following
properties:

\begin{enumerate}
\item dS-invariance:
\[ {\cal W}^{(1/2)}(\ksi{1}_{g},\ksi{2}_{g})=
U_s (g){\cal W}^{(1/2)}(\ksi{1},\ksi{2})\overline{U}_s (g)\]
where $U_s (g)$ is the identical representation at $s=0$ and the four-spinor
representation at $s=1/2$.

\item Causality:
\[ {\cal W}^{(s)}(\stack{1}{x},\stack{2}{x})=
{\cal W}^{(s)}(\stack{2}{x},\stack{1}{x}) \qquad
\stack{1}{x}_A \stack{2}{x}^A >-R^2.\]

\item Regularized function ${\cal W}^{(s)}(\stack{1}{x},\stack{2}{x})$
is the boundary value of the function
${\cal W}^{(s)}(\stack{1}{\zeta},\stack{2}{\zeta})$
which is analytic in  certain domain of the complexified dS space.
\end{enumerate}
For the spin zero case the above properties were proved in~\cite{24}; but
also in this case the CS method gives the sufficient simplification
since the property~1 is found almost obvious. In the theorem~4.1 of the
mentioned paper the property of {\it positive definiteness}
\[\int \int \frac{\d^4 \stack{1}{x}}{\stack{1}{x}^5}
\frac{\d^4 \stack{2}{x}}{\stack{2}{x}^5}
{\cal W}^{(0)}(\stack{1}{x},\stack{2}{x}) f(\stack{1}{x})
f^* (\stack{2}{x})  >  0\]
was proved for any functions $f(x)$ such that this integral is meaningful,
where ${\cal W}^{(0)}(\stack{1}{x},\stack{2}{x})$
is considered in the sense of boundary values. For the spin~1/2 case this
property may be proved in the completely analogous way since
\[\int \int \frac{\d^4 \stack{1}{x}}{\stack{1}{x}^5}
\frac{\d^4 \stack{2}{x}}{\stack{2}{x}^5} \overline{\psi}(\stack{1}{x})
{\cal W}^{(1/2)}(\stack{1}{x},\stack{2}{x}) \psi(\stack{2}{x})  =
\int_{{\Bbb R}^3}\d^3 w \, \left| \int\frac{\d^4 x}{x^5} \overline{\psi}(x)
\Phi_{{\bi w}}^{(1/2)-}(x)\right|^2 .\]
Defining the creation-annihilation operators so that they
possess the necessary commutation relations, we can construct
 quantized fields $\phi^{(s)}(x)$;
the propagators of these fields are equal to
\[ [\phi^{(s)}(\stack{1}{x}),
\overline{\phi}\vphantom{\phi}^{(s)}(\stack{2}{x})
]_\pm ={\cal W}^{(s)}(\stack{1}{x},\stack{2}{x})-
{\cal W}^{(s)}(\stack{2}{x},\stack{1}{x}) \]
and therefore are dS-invariant and causal automatically.

\ack

I am grateful to Yu P Stepanovsky for the constant support during
the fork and to  Ph Spindel sending me a copy of his paper~\cite{80}.

\end{document}